\documentclass[aps,preprint,nofootinbib,superscriptaddress]{revtex4}

\usepackage{amssymb}

\usepackage{epsfig}
\usepackage[bookmarksnumbered,bookmarksopen,colorlinks,citecolor=blue,linkcolor=blue]{hyperref}

\begin{document}

\title{Dynamical nucleus-nucleus potential and incompressibility of nuclear matter }

\author{V. Zanganeh}
\affiliation{ Department of Physics, Guangxi Normal University,
Guilin 541004, People's Republic of China } \affiliation{
Department of Physics, University of Mazandaran, Babolsar 47415,
Iran}

\author{N. Wang}
\email{wangning@gxnu.edu.cn}\affiliation{ Department of Physics,
Guangxi Normal University, Guilin 541004, People's Republic of
China }

\author{O. N. Ghodsi}
\affiliation{ Department of Physics, University of Mazandaran,
Babolsar 47415, Iran}

\begin{abstract}

The dynamical nucleus-nucleus potentials for some fusion reactions
are investigated by using the improved quantum molecular dynamics
(ImQMD) model with different sets of parameters in which the
corresponding incompressibility coefficient of nuclear matter is
different. Two new sets of parameters  SKP* and IQ3  for the ImQMD
model are proposed with the incompressibility coefficient of 195
and 225 MeV, respectively. The measured fusion excitation function
for $^{16}$O+$^{208}$Pb and the charge distribution of fragments
for Ca+Ca and Au+Au in multi-fragmentation process can be
reasonably well reproduced. Simultaneously, the influence of the
nuclear matter incompressibility and the range of nucleon-nucleon
interaction on the nucleus-nucleus dynamic potential is
investigated.

\end{abstract}
\maketitle

\begin{center}
\textbf{I. INTRODUCTION}
\end{center}

The synthesis of super-heavy nuclei and heavy-ion fusion at deep
sub-barrier energies have attracted a great deal of attention in
recent years \cite{Hof00,Ogan10,Zag08,Ada,Zhou11,Wang11,Sar,Timm}.
The calculation of nucleus-nucleus potential especially at short
distances is of crucial importance for these studies. Several
static and dynamic models have been proposed for calculating the
nucleus-nucleus potential \cite{prox,liumin06,MWS,monte
carlo,TDHF,Umar12,ImQMD2010}. The static potentials are usually
obtained by some empirical formulas or based on the double-folding
concept and the sudden approximation. To consider the dynamical
process in fusion reaction microscopically, some microscopic
dynamics models, such as the time-dependent Hartree-Fock (TDHF)
model \cite{TDHF,Umar12} and the improved quantum molecular
dynamic (ImQMD) model \cite{ImQMD2002,ImQMD2004} have been
developed. The ImQMD model is a semi-classical microscopic
dynamics model and is successfully applied on intermediate-energy
heavy-ion collisions and heavy-ion reactions at energies around
the Coulomb barrier \cite{ImQMD2004,ImQMD05,ImQMD2010}. In
Ref.\cite{ImQMD2010} the extended Thomas-Fermi approximation is
adopted for calculation of the dynamical nucleus-nucleus potential
at short distances based on the obtained dynamical densities of
the reaction system from the ImQMD simulations. The energy
dependence of the dynamical nucleus-nucleus potential was
observed. In addition to the influence of the incident energy on
the nucleus-nucleus potential due to the dynamical evolution of
the densities, the influence of the  incompressibility coefficient
of nuclear matter on the nucleus-nucleus potential and the fusion
cross sections are also investigated in Refs. \cite{esbensen,nucl}
according to the double-folding calculation with the M3Y
interactions.

To investigate the influence of nuclear equation of state on the
dynamical nucleus-nucleus potential, we will study the
nucleus-nucleus potential at short distances with the ImQMD model
by adopting different sets of parameters. The corresponding
nuclear equation of state for these different sets of parameters
is different, with which we attempt to understand the influence of
the incompressibility coefficient of nuclear matter on the
dynamical fusion potential. The structure of this paper is as
follows: In sec. II, the ImQMD model is briefly introduced. In
sec. III, two sets of parameters IQ3 and SKP* are proposed for the
ImQMD calculation, and the fusion reactions $^{16}$O+$^{208}$Pb
and $^{48}$Ca+$^{208}$Pb at energies around the Coulomb barrier
and the reactions Ca+Ca and Au+Au at incident energy of
35MeV/nucleon are also studied for testing IQ3 and SKP*. In
addition, the dynamical nucleus-nucleus potential inside the
Coulomb barrier is simultaneously investigated. Finally the
conclusion is given in Sec. IV.

\begin{center}
\noindent{\bf {II. THE IMPROVED QUANTUM MOLECULAR DYNAMICS MODEL}}\\
\end{center}

In the ImQMD model, as in the original QMD model \cite{QMD},  each
nucleon is represented by a coherent state of a Gaussian wave
packet. The density distribution function $\rho$ of a system reads
\begin{equation} \label{1}
\rho(\mathbf{r})=\sum_i{\frac{1}{(2\pi \sigma_r^2)^{3/2}}\exp
\left [-\frac{(\mathbf{r}-\mathbf{r}_i)^2}{2\sigma_r^2} \right ]},
\end{equation}
where $\sigma_r$ represents the spatial spread of the wave packet.
The propagation of nucleons is governed by Hamiltonian equations
of motion under the self-consistently generated mean field,
\begin{equation} \label{2}
\mathbf{\dot{r}}_i=\frac{\partial H}{\partial \mathbf{p}_i}, \; \;
\mathbf{\dot{p}}_i=-\frac{\partial H}{\partial \mathbf{r}_i},
\end{equation}
where $r_i$ and $p_i$ are the center of the $i$-th wave packet in
the coordinate and momentum space, respectively.  The Hamiltonian
$H$ consists of the kinetic energy
$T=\sum\limits_{i}\frac{\mathbf{p}_{i}^{2}}{2m}$ and the effective
interaction potential energy $U$:
\begin{equation} \label{3}
H=T+U.
\end{equation}
The effective interaction potential energy is written as the sum
of the nuclear interaction potential energy $U_{\rm
loc}=\int{V_{\rm loc}(\textbf{r})d\textbf{r}}$ and the Coulomb
interaction potential energy $U_{\rm Coul}$ which includes the
contribution of the direct and exchange terms,
\begin{equation}
U=U_{\rm loc}+U_{\rm Coul}.
\end{equation}
Where $V_{\rm loc}(r)$ is the potential energy density that is
obtained from the effective Skyrme interaction and taken to be the
same as that in Ref. \cite{ImQMD2004}:
\begin{equation}
V_{\rm
loc}=\frac{\alpha}{2}\frac{\rho^2}{\rho_0}+\frac{\beta}{\gamma+1}\frac{\rho^{\gamma+1}}{\rho_0^{\gamma}}+\frac{g_{\rm
sur}}{2\rho_0}(\nabla\rho)^2
+g_{\tau}\frac{\rho^{\eta+1}}{\rho_0^{\eta}}+\frac{C_s}{2\rho_0}[\rho^2-k_s(\nabla\rho)^2]\delta^2
\end{equation}
where $\delta=(\rho_n -\rho_p)/(\rho_n +\rho_p)$ is the isospin
asymmetry. To describe the fermionic nature of the N-body system
and to improve the stability of an individual nucleus, the
phase-space occupation constraint method \cite{constrain} and the
system-size-dependent wave-packet width $\sigma_r =  \sigma_0 +
\sigma_1 A^{1/3} $ fm \cite{ImQMD2002} are adopted. The parameter
sets adopted in this work are shown in Table I.

\newpage
\begin{table}
 \caption{ Model parameters adopted in this work.}
\begin{tabular}{lccccccccccc}
\hline Parameter & $\alpha $ & $\beta $ & $\gamma $ &$%
g_{\rm sur}$ & $ g_{\tau }$ & $\eta $ & $C_{s}$ & $\kappa _{s}$ &
$\rho
_{0}$ & ~~$\sigma_0$~~ & ~~$\sigma_1$~~ \\
 & (MeV) & (MeV) &  & (MeVfm$^{2}$) & (MeV) &  & (MeV) & (fm$^{2}$) &
 (fm$^{-3}$) & (fm) & (fm) \\ \hline
IQ2 & -356 & 303 & 7/6 & 7.0 & 12.5 & 2/3 & 32.0 & 0.08 & 0.165 & 0.88 & 0.09\\
SKP*& -356 & 303 & 7/6 & 19.5 & 13.0 & 2/3 & 35.0 & 0.65 & 0.162 &0.94 & 0.018\\
IQ3 & -207 & 138 & 7/6 & 18.0 & 14.0 & 5/3 & 32.0 & 0.08 & 0.165 & 0.94 & 0.018\\
  \hline
\end{tabular}
\end{table}

\begin{center}
\textbf{III. RESULTS}
\end{center}

In this section we first briefly introduce the parameter sets IQ3
and SKP*. Then we test the new parameter sets through fusion
reactions and heavy-ion collisions at intermediate energy. Finally
we investigate the dynamical nucleus-nucleus potential inside of
the Coulomb barrier.

\begin{center}
\textbf{A. New parameter sets SKP* and IQ3  }
\end{center}

According to the properties of nuclei at ground state and the
knowledge of nuclear incompressibility, a number of Skyrme forces,
such as SkM* \cite{skm}, SKP \cite{skp} and SLy4 \cite{sly4} were
proposed in recent decades. With the proposed parameter sets, the
Skyrme energy-density functionals have been successfully applied on
the studies of nuclear structure, fusion reaction and neutron star,
etc. Based on the parameters of Skyrme forces, the ImQMD parameters
can be directly obtained as those done in Ref. \cite{ImQMD05}.
Considering the range of nucleon-nucleon interaction in the ImQMD
model which is represented by a gaussian wave-packet, the parameters
of the ImQMD model in this work are re-adjusted for studying the
fusion reactions in which the stability of an individual nucleus
plays a role for a reliable description of the dynamical process.

The knowledge of nuclear equation of state at densities around the
normal density $\rho_0$ is helpful to constrain the model
parameters. In the ImQMD model, the incompressibility coefficient
$K_{\infty }$ of symmetric nuclear matter at $\rho_0$ is expressed
as
\begin{equation}
K_{\infty } = 9\rho _{0}^{2}  \frac{\partial ^{2}(E/A)}{\partial \rho ^{2}}%
{\Bigg |} _{\rho=\rho _{0}} =-2\xi c_k\rho_0^{2/3}+9\beta (\gamma
-1)\gamma +9g_{\tau }(\eta -1)\eta
\end{equation}%
with
$c_k=\frac{\hbar^2}{2m}\frac{3}{5}(\frac{3\pi^2}{2})^{2/3}=75.0$
MeV fm$^2$.  $\xi=c_0/c_k$ is a correction factor for the kinetic
energy of a nuclear system when applying the extended Thomas-Fermi
(ETF) approximation which is roughly expressed as
\begin{equation}
E_k^{\rm ETF}\simeq
c_{0}\sum_i{\rho_i^{2/3}}+\frac{c_{1}}{\sum{\rho_i}}\sum_{i,j\neq
i}{f_s \rho_{ij}}+c_{2} N
\end{equation}
in this model. The expressions of $\rho_i$ and $\rho_{ij}$ are given
in Ref. \cite{ImQMD2010}. The coefficients $c_{0}$, $c_{1}$ and
$c_{2}$ can be determined by the kinetic energies $T$ of nuclei at
their ground state. $\xi=1$ is for the idealistic Fermi-gas.
Considering the  range of realistic nucleon-nucleon interactions
which is described by gaussian wave-packets in the ImQMD model, we
approximately set $\xi=0.4\sim0.6$ for a reasonable description of
the  properties of nuclei at ground state and the stability of an
individual nucleus. The corresponding energy per particle $E/A$ at
$\rho_0$ reads
\begin{equation}
 \frac{E(\rho)}{A} {\Bigg |} _{\rho=\rho _{0}} = \xi c_k \rho _{0}^{2/3} +\frac{\alpha }{2}+\frac{\beta }{\gamma +1}%
+g_{\tau }.
\end{equation}%
Although experimental and theoretical investigations on the nuclear
equation of state suggest that $K_{\infty}\approx 230 $ MeV,
$E/A\approx -16$ MeV around the saturation density which is $\sim
0.16$ fm$^{-3}$ \cite{incom1,incom2,incom3}, the uncertainty of
nuclear equation of state still causes some difficulties for a
unambiguous determination of the model parameters.

To investigate the influence of the nuclear matter
incompressibility coefficient on fusion reactions, we attempt to
propose two new sets of parameters, SKP* and IQ3 for the ImQMD
simulations. The parameter set SKP* is generally determined based
on the Skyrme force SKP \cite{skp} in which the parameters
$g_\tau$, $\sigma_0$ and $\sigma_1$ are adjusted for an
appropriate description of nuclear properties at ground state and
the fusion reactions. For both IQ2 and SKP*, the corresponding
values of $K_{\infty}$ are the same but the wave-packet widths
$\sigma_r$ are different, which is useful for exploring the
influence of the interaction range of nucleons and the finite-size
effect of nuclei. To explore the influence of the nuclear matter
incompressibility on the fusion reactions, we also construct the
parameter set IQ3 in which the wave-packet widths $\sigma_r$ are
the same as those in SKP* but the incompressibility coefficient
$K_{\infty} = 225$ MeV is obviously larger than that in SKP* and
IQ2 (see Table II). The corresponding kinetic energy coefficients
for different sets of parameters are listed in Table II.

\begin{table}
\caption{Kinetic energy coefficients in the ETF approximation and
incompressibility coefficient for parameter sets IQ2, SKP* and
IQ3.}
\begin{tabular}{l c c c c }
 \hline
Parameter   &\qquad $c_{0}$ (MeV $fm^2$)   & \qquad $c_{1}$ (MeV $fm^2$)  &\qquad
$c_{2}$ (MeV) & ~~~~$K_{\infty}$ (MeV) \\  \hline
IQ2                 &\qquad 41.2       &\qquad 4.8         &\qquad $-1.0$   & 195  \\
SKP*                &\qquad 38.9       &\qquad 0.1         &\qquad  0   & 195  \\
IQ3                 &\qquad 43.3       &\qquad 0.34        &\qquad  0    & 225  \\
\hline
\end{tabular}
\end{table}

\begin{center}
\textbf{B. Tests for IQ3 and SKP*}
\end{center}

With the parameter sets IQ3 and SKP*, the time evolutions of the
binding energies and nuclear radii for a number of nuclei have
been checked. We find that an individual nucleus can remain stable
for several thousands fm/c without spurious nucleon emission.
Simultaneously, some fusion reactions are investigated for testing
the parameter sets. Fig. 1 shows the time evolution of the
densities for the fusion reaction $^{48}$Ca+$^{208}$Pb at an
incident energy of $E_{\rm c.m.}=200$ MeV with the parameter set
IQ3. From the time evolution, one sees that the central densities
of nuclei in the reactions are reasonable. In addition, we note
that the surface diffuseness of nuclei at neck region increases
when the neck of the di-nuclear system is well formed.

\begin{figure}
\includegraphics[angle=0,width=1.0 \textwidth]{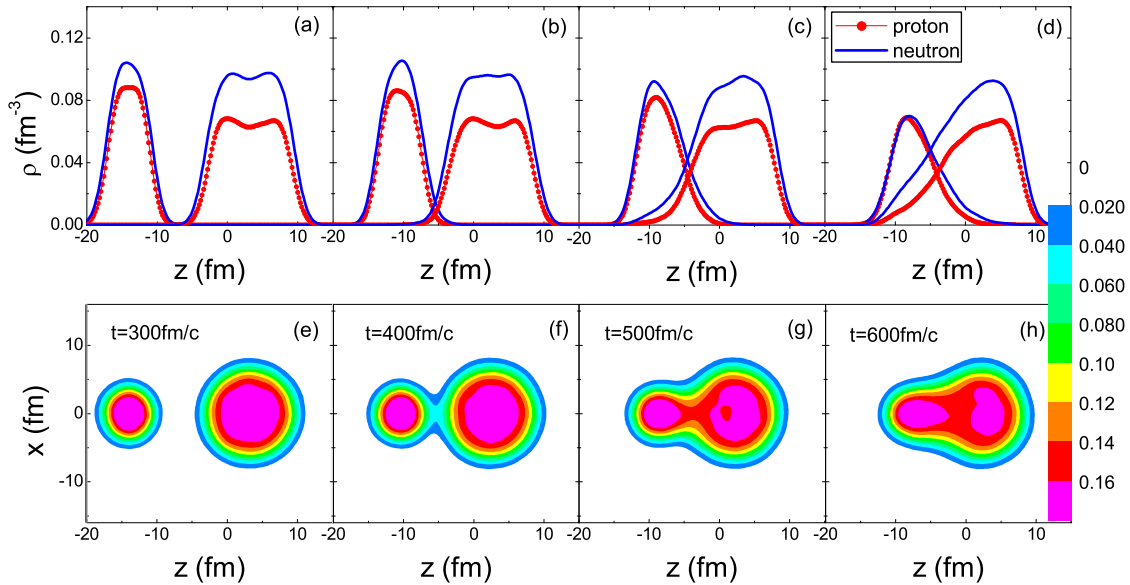}
\caption{(Color online) Time evolution of the density distribution
for the fusion reaction $^{48}$Ca+$^{208}$Pb at an incident energy
of $E_{\rm c.m.}=200$ MeV with IQ3. }
\end{figure}

The fusion cross sections of $^{16}$O+$^{208}$Pb are also calculated
with the ImQMD model by adopting IQ3 and SKP*. Through creating
certain bombarding events (about 100) at each incident energy
$E_{\rm c.m.}$ and at each impact parameter $b$, and counting the
number of fusion events, we obtain the fusion probability $g_{\rm
fus}(E_{\rm c.m.},b)$ of the reaction, by which the fusion cross
section can be calculated \cite{ImQMD2002}:
\begin{equation}
\sigma _{\rm fus}(E_{\rm c.m.})=2\pi \int b \, g_{\rm fus} \, db
\simeq 2\pi \sum b \, g_{\rm fus} \, \Delta b.
\end{equation}
The initial distance between the projectile and target is taken to
be $R=40$ fm for calculating the fusion cross sections. Fig. 2 shows
the comparison of our calculated results and the experimental data
for the fusion reaction $^{16}$O+$^{208}$Pb. The solid and open
circles denote the experimental data and the calculation results,
respectively. The measured fusion excitation function for
$^{16}$O+$^{208}$Pb can be reasonably well reproduced with the new
parameter set IQ3 and SKP* at energies near and above the Coulomb
barrier. The over-prediction of the fusion cross sections at
sub-barrier energies is due to the shell effect of doubly-magic
nuclei that is not well described with this semi-classical model.

\begin{figure}
\includegraphics[angle=0,width=0.9\textwidth]{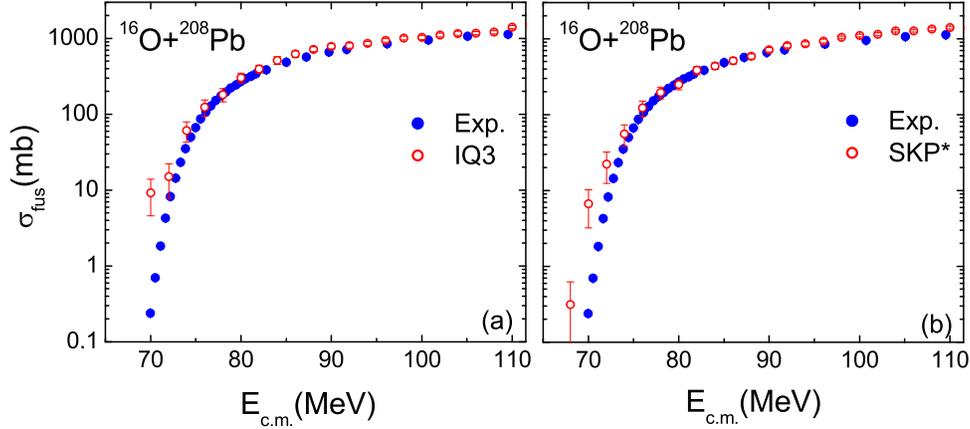}
\caption{(Color online) Fusion excitation function of
$^{16}$O+$^{208}$Pb reaction. The solid circles denote the
experimental data \cite{Mort99}. The open circles denote the results
of ImQMD with parameter sets IQ3 and SKP*. }
\end{figure}

For further testing the reliability of IQ3 and SKP*, we have also
calculated the charge distributions of fragments in
multi-fragmentation processes at intermediate energy heavy-ion
collisions. In Fig. 3 we show the charge distribution of fragments
by using the ImQMD model with parameters set of IQ3 and SKP* for
$^{40}$Ca+$^{40}$Ca \cite{expcaca} and $^{197}$Au+$^{197}$Au
\cite{expAuAu} at incident energy of 35 MeV/nucleon. Here we
create 500 events for head-on collisions and for each event we
self-consistently simulate the whole collision process till $t =
6000$ fm/c with a step size of $\Delta t=1$ fm/c. We have found
that the experimental data can be reproduced remarkably well.

\begin{figure}
\includegraphics[angle=0,width=0.8 \textwidth]{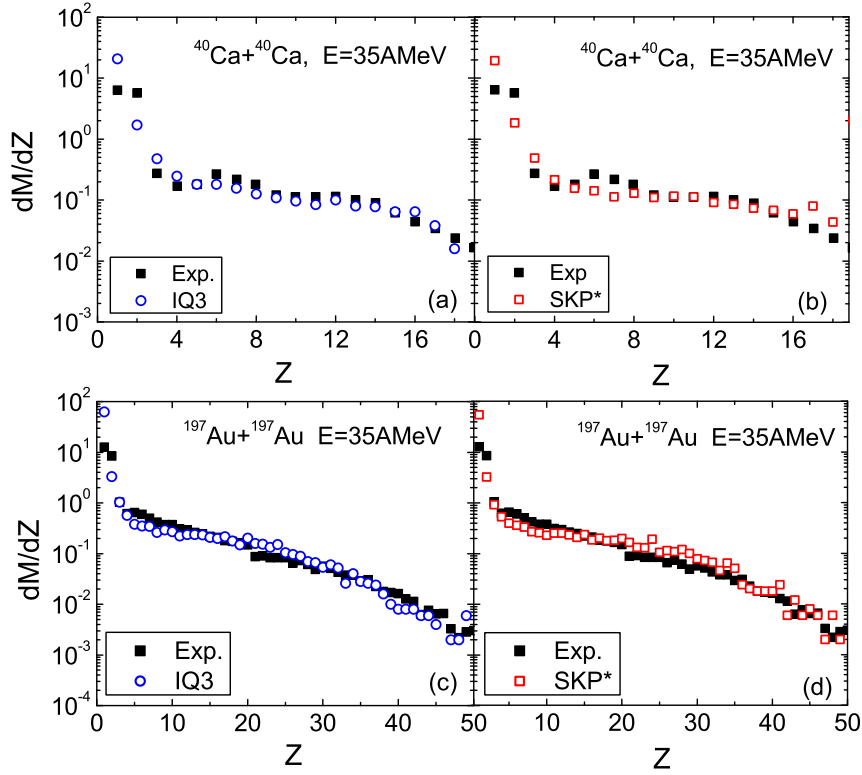}
\caption{(Color online) Charge distribution of fragments for central
collision of $^{197}$Au+$^{197}$Au and $^{40}$Ca+$^{40}$Ca at
35AMeV. The solid squares denote the experimental data taken from
Refs.\cite{expcaca,expAuAu}. The open circles and open squares
denote the results of ImQMD with IQ3 and SKP*, respectively. }
\end{figure}

\begin{figure}
\includegraphics[angle=0,width=0.9 \textwidth]{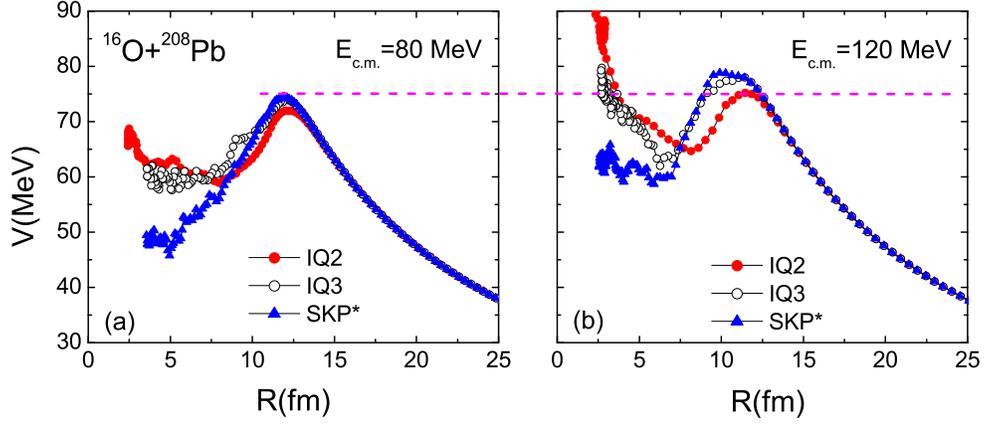}
\caption{(Color online) Dynamical nucleus-nucleus potential for
the reaction $^{16}$O+$^{208}$Pb  with parameter set IQ2, IQ3 and
SKP* at incident energies of (a) $E_{\rm c.m.}=80$ MeV  and (b)
$E_{\rm c.m.}=120$ MeV, respectively. Here we set the initial
distance between two nuclei as 30 fm. The solid circles, open
circles and triangles denote the results with IQ2, IQ3 and SKP*,
respectively. The dashed line is to guide the eyes.}
\end{figure}

\begin{figure}
\includegraphics[angle=0,width=0.9 \textwidth]{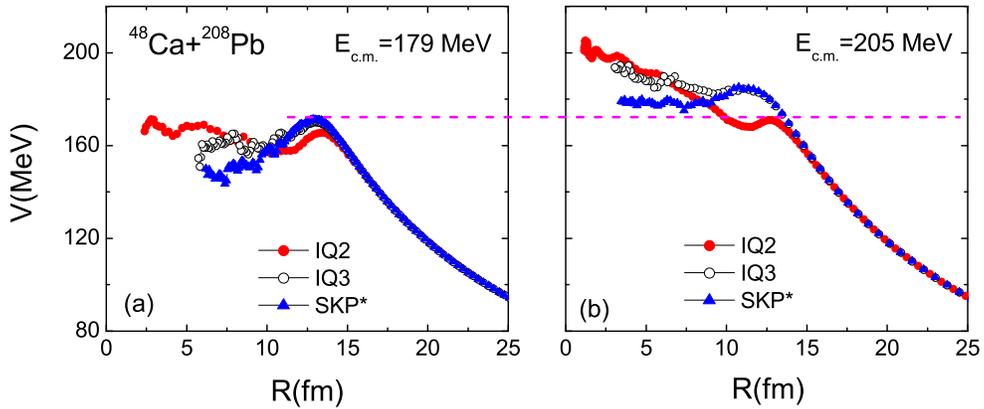}
\caption{(Color online) The same as Fig. 4, but for the reaction
$^{48}$Ca+$^{208}$Pb at incident energies of (a) $E_{\rm c.m.}=179
$ MeV and (b) $E_{\rm c.m.}=205$ MeV, respectively.  }
\end{figure}

\begin{center}
\textbf{C. Dynamical nucleus-nucleus potential}
\end{center}

Based on the dynamical densities of the reaction system, the
nucleus-nucleus potential can be obtained with the ETF
approximation for the kinetic energies \cite{ImQMD2010}. After the
di-nuclear system is formed, the nucleus-nucleus potential may be
described by a way like the entrance channel potential
\cite{liumin06}
\begin{equation}
 V (R) =E_{\rm tot}(R)-\bar E_{1}-\bar E_{2},
\end{equation}
where $E_{\rm tot}(R)$ is the energy of the composite system which
is strongly dependent on the dynamical density distribution of the
system obtained with the ImQMD model, $\bar E_{1}$ and $\bar E_{2}$
are the time average of the energies of the projectile and target
nuclei, respectively. In this work, the dynamical nucleus-nucleus
potential are calculated the same as in Ref. \cite{ImQMD2010}, but
with the parameter sets IQ3 and SKP*. The kinetic energy
coefficients $c_0$, $c_1$ and $c_2$ listed in Table II are
determined by fitting the obtained kinetic energies $T$ of a series
of nuclei from light to heavy nuclei at their ground state with
Eq.(7). To investigate the dynamical nucleus-nucleus potential, the
head-on collisions of $^{16}$O+$^{208}$Pb and $^{48}$Ca+$^{208}$Pb
at two different incident energies with three different parameter
sets have been studied. As mentioned previously, the values of the
incompressibility coefficient are the same but the wave-packet
widths are different for IQ2 and SKP*, while the incompressibility
coefficients are different but the wave-packet widths are the same
for IQ3 and SKP*.

Fig. 4 and Fig. 5 show the dynamical nucleus-nucleus potential for
$^{16}$O+$^{208}$Pb and $^{48}$Ca+$^{208}$Pb  with the parameter
sets IQ2, IQ3 and SKP*, respectively. From the figures we find that:
(1) the dynamical barrier height depends on the incident energy as
mentioned in Ref. \cite{ImQMD2010}; (2) the wave-packet width
influences nuclear surface diffuseness and thus influences both the
potential barrier height and the potentials at short distances; and
(3) the nuclear matter incompressibility seems just to affect the
potentials at short distances if taking the same wave-packet width.
Comparing the results with SKP* and IQ3, one sees that the
potentials at short distances increase with the increase of the
incompressibility coefficient. To illustrate this point, we also
study the static entrance channel potential \cite{liumin06} of
$^{48}$Ca+$^{208}$Pb with the Skyrme energy-density function by
adopting different parameter sets. Fig. 6 shows the nuclear
potential (i.e., removing the Coulomb potential from the entrance
channel nucleus-nucleus potential) as a function of distance between
two nuclei. Obviously, the nuclear potentials do increase with the
value of $K_{\infty}$. In addition, the obtained dynamical
nucleus-nucleus potential has been checked by directly using the
barrier penetration calculations for the fusion cross section. As an
example, the fusion cross section for the reaction
$^{16}$O+$^{208}$Pb at $E_{\rm c.m.}=80$ MeV is calculated based on
the obtained potential for head-on collisions with IQ3 which is
shown in Fig.4(a). We find that the obtained fusion cross section
with the barrier penetration approach is very close to the result
with Eq.(9). We also note that the obtained distribution function
for the fusion probability $g_{\rm fus}(b)$ is different with the
two approaches for this reaction, although the calculated fusion
cross section is close to each other. The difference is due to that
the reduced mass $\mu$ in the traditional barrier penetration
calculations is fixed, but it changes as a function of distance $R$
between two nuclei in the ImQMD simulations when the neck of the
composite system is formed.

\begin{figure}
\includegraphics[angle=0,width=0.7\textwidth]{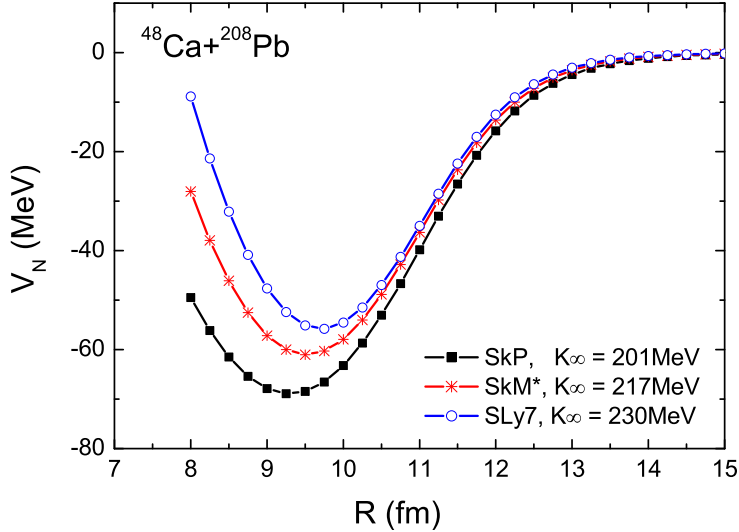}
\caption{(Color online) Nuclear potential of $^{48}$Ca+$^{208}$Pb
with different Skyrme forces. The solid squares, stars and open
circles denote the results with SKP, SkM* and SLy7, respectively.
}
\end{figure}

\begin{figure}
\includegraphics[angle=0,width=1\textwidth]{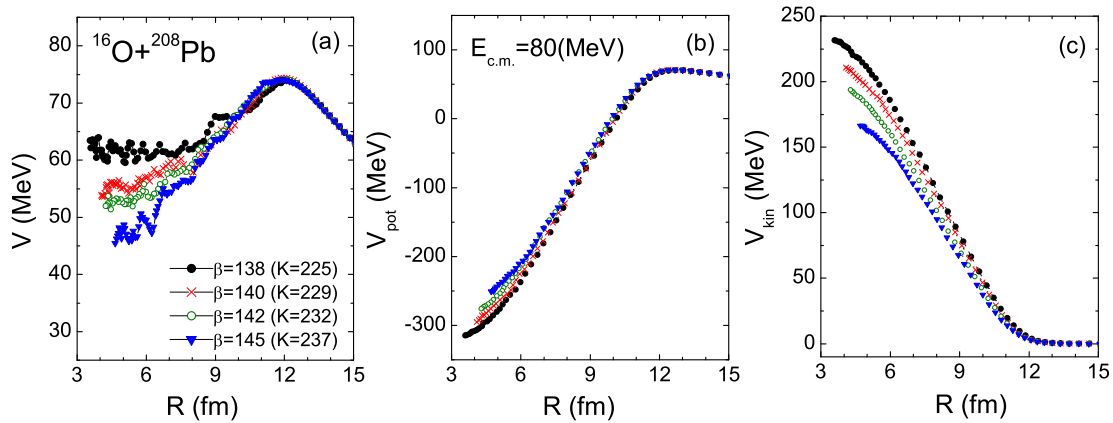}
\caption{(Color online) (a) Dynamical nucleus-nucleus potential,
(b) contribution of the interaction potential energy and (c)
contribution of the kinetics energy for reaction
$^{16}$O+$^{208}$Pb at $E_{\rm c.m.}=80$ MeV based on the
parameter set IQ3 but varying the value of $\beta$. The solid
circles, crosses, open circles and triangles denote the results
with $\beta$=138, 140, 142 and 145 MeV, respectively.}
\end{figure}

To further investigate the influence of nuclear repulsion on the
nucleus-nucleus potential, we study the nucleus-nucleus potential
with IQ3 but varying the parameter $\beta$. The value of $\beta$
and the corresponding kinetic energy coefficients and $K_{\infty}$
are listed in Table III (here $c_3=0$). Fig. 7 shows the dynamical
nucleus-nucleus potential and the contribution of the
corresponding effective interaction potential energy and that of
the corresponding kinetics energy for the reaction
$^{16}$O+$^{208}$Pb with different values of $\beta$. One can see
from Fig. 7 (b) that the contribution of the effective interaction
potential energy increases with the nuclear incompressibility
coefficient as we expected. But we also note that the change of
$\beta$ causes larger change of the corresponding kinetic energy
than that of the interaction potential energy, which results in
the decrease of the total nucleus-nucleus potential at short
distances with increasing
 the nuclear repulsion.

\begin{table}
\caption{The same as Table II, but taking different value for
$\beta$.}
\begin{tabular}{c c c c}
\hline
Parameter        & ~~~~$c_{0}$ (MeV $fm^2$) ~~~~  &  ~~~~ $c_{1}$ (MeV $fm^2$) ~~~~  & $K_{\infty} $ (MeV) \\
\hline
$\beta=138$       &  43.3        & 0.34          &  225  \\
$\beta=140$       &  40.8       &  0.30         &  229  \\
$\beta=142$       &  38.0       &  0.83         &  232  \\
$\beta=145$       &  33.8       &  1.43         &  237  \\
\hline \end{tabular}
\end{table}

\begin{center}
\textbf{IV. CONCLUSION}
\end{center}

In this work, the dynamical nucleus-nucleus potentials for fusion
reactions have been investigated by using the improved quantum
molecular dynamics model with different sets of parameters. By using
two new sets of parameters IQ3 and SKP* with which the measured
fusion excitation function for $^{16}$O+$^{208}$Pb and the charge
distribution of fragments for Ca+Ca and Au+Au in multi-fragmentation
process can be reasonably well reproduced, we find that both the
nuclear incompressibility and the range of nucleon-nucleon
interactions significantly influence the nucleus-nucleus potential.
The interaction range represented by the gaussian wave-packet width
affects both the barrier height and the potentials at short
distances. The incompressibility coefficient of nuclear matter
mainly influences the potentials at short distances if taking the
same wave-packet width. In addition, the nuclear repulsion
influences both the effective interaction potential energy and the
kinetic energy of a fusion system.

\begin{center}
\textbf{ACKNOWLEDGEMENTS}
\end{center}
This work was supported by National Natural Science Foundation of
China, Nos 10875031, 11005003 and 10979024. One of the authors (V.
Z.) acknowledges support from the General Department of
Scholarships and Student Affairs, Ministry of Science and
Technology, Iran.

\end{document}